\documentclass[aps,prl,twocolumn,superscriptaddress,notitlepage]{revtex4-1}
\usepackage[final]{graphicx}
\usepackage{epstopdf}
\begin{document}

\title{Discrete Electronic Sub-bands without Confinement: Bragg Scattering at Molecular Edges}

\author{A. Mart\'in-Jim\'enez}
\affiliation{Instituto Madrile\~no de Estudios Avanzados en Nanociencia (IMDEA-NANO), Madrid, Spain}
\author{J. M. Gallego}
\affiliation{Instituto de Ciencia de Materiales de Madrid (ICMM-CSIC), Madrid, Spain}
\author{R. Miranda}
\affiliation{Dep. de F\'isica de la Materia Condensada, Universidad Aut\'onoma de Madrid, 28049 Madrid, Spain}
\affiliation{Instituto Madrile\~no de Estudios Avanzados en Nanociencia (IMDEA-NANO), Madrid, Spain}
\author{R. Otero}
\affiliation{Dep. de F\'isica de la Materia Condensada, Universidad Aut\'onoma de Madrid, 28049 Madrid, Spain}
\affiliation{Instituto Madrile\~no de Estudios Avanzados en Nanociencia (IMDEA-NANO), Madrid, Spain}

\email[]{roberto.otero@uam.es}
\date{\today}

\begin{abstract}
The discretization of the electronic structure of nanometer-size solid systems due to quantum confinement and the concomitant modification of their physical properties is one of the cornerstones for the development of Nanoscience and Nanotechnology. In this letter we demonstrate that Bragg scattering of Cu(111) surface state electrons by the periodic arrangement of tetracyanoquinodimethane (TCNQ) molecules at the edges of self-assembled molecular islands, discretizes the possible values of the electron momentum parallel to the island edge. The electronic structure consists thus of a discrete number of sub-bands which occur in a non-closed space and, therefore, without quantum confinement.
\end{abstract}


\maketitle

The discretization of the electronic energy levels upon confinement in regions smaller than their coherence length is a fundamental result of Quantum Mechanics \cite{cohen1977} that determines the electronic structure of solid state nanostructures. It is a key ingredient to understand phenomena as diverse as the optical and transport properties of semiconductor Quantum Dots, Wires and Wells \cite{harrison2005}, the oscillatory behavior of the superconducting transition temperature in thin films \cite{Qin2009}, the thermal stability of metallic thin films \cite{Calleja2006,Otero2002,Su2001,Yeh2000}, or the magnetic coupling across thin non-magnetic spacers \cite{Cebollada1991,Ortega1992} to mention just a few examples. In solid state nanostructures, confinement is usually achieved by reflection of the electronic wave function at their surfaces or interfaces with different materials. When the width of the material between two such interfaces is small compared to the electronic coherence length, the values of the perpendicular wave vector can only take on  a series of size-dependent discrete values, $k_{\perp,n}$. If the electronic structure of the solid in bulk form is characterized by a dispersion relation $E(\mathbf{k_{\parallel}},k_{\perp})$, that of the nanostructure will consist in a discrete number of subbands for the discrete series of $k_{\perp,n}$ values, i.e. $E_{n}(\mathbf{k_{\parallel}},k_{\perp,n})$. In the absence of any other discretization mechanism, one single confining interface is not enough to obtain a discrete sub-band structure: reflection at the interface would simply preserve $k_{\parallel}$ and reverse $k_{\perp}$, and the interference between incoming and outgoing electrons would lead to electron standing waves with a continuous range of momenta.

\begin{figure}[b]
\includegraphics[]{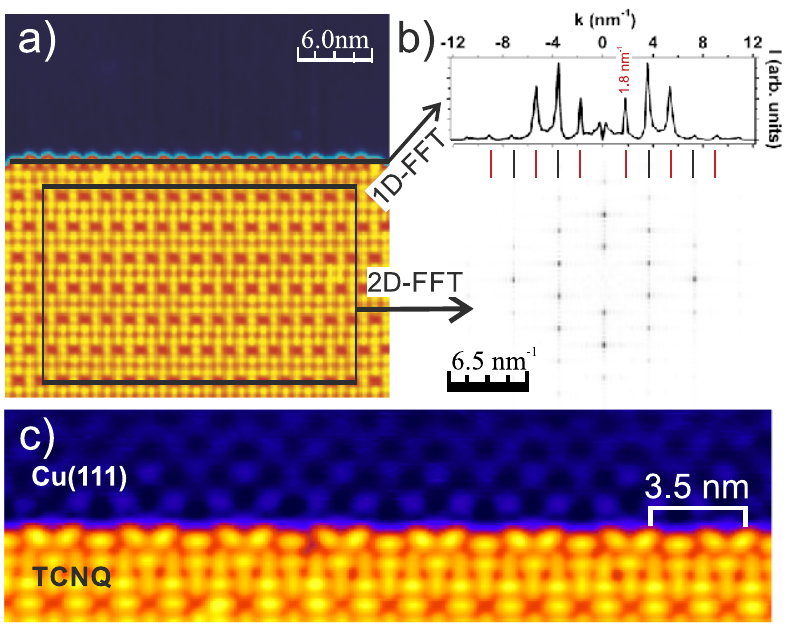}
\caption[]{
\label{figure1}
(a) STM image of the edge of a TCNQ island, showing the bulk and edge molecular arrangements. (b) 2D FFT  of the island bulk (lower panel) and 1D FFT of a scan line over the island edge (upper panel). Black tick-marks correspond to the peaks in the 1D FFT that correspond to the projection of the bulk reciprocal space over the edge direction, while red ticks mark the position of the new peaks. (c) STM image recorded at low voltage where the standing wave pattern is visible.
}
\end{figure}

In this letter we show Low-Temperature Scanning Tunnelling Microscopy and Spectroscopy results demonstrating that the electronic structure of the two-dimensional electron gas (2DEG) at the Cu(111) surface nearby the edges of self-assembled TCNQ islands consists of a discrete set of sub-bands corresponding to a discrete set of $k_{\parallel,n}$ values even when other confining interfaces are much farther apart than the electron's coherence length. The values of $k_{\parallel,n}$ are equally spaced by an amount $G/2$ where $G=2\pi/a$ is the reciprocal lattice vector corresponding to the periodicity $a$ of the molecular edge. Our analysis reveals that this effect arises from coherent Bragg diffraction of surface state electrons in backscattering configuration. To our knowledge, this is the first example in which the discretization of the electron momentum and electronic structure around solid state nanostructures is not caused by electron confinement but by diffraction, leading to a 2D periodic perturbation of the charge density and Density of States of the surface surrounding molecular islands with important implications for the adsorption and reaction of further adsorbates.

The experiments were carried out in an Ultra-High Vacuum (UHV) chamber ($P\sim10^{-10}$ Torr) equipped with an Omicron Low-Temperature Scanning Tunneling Microscope (LT-STM) and facilities for sample cleaning and evaporation of organic material. Clean Cu(111) surfaces with large terraces were obtained by sputtering the sample with Ar$^{\mathrm{+}}$ ions ($P_{Ar}\sim10^{-5}$  Torr, 1.5 keV) followed by annealing to 500 K for 10 minutes. TCNQ molecules were deposited from a molecular evaporator (sublimation temperature 360 K) with the substrate held at RT. $dI/dV(E)$ were recorded with a lock-in amplifier using a modulation voltage of 10-20 mV under open feedback conditions. For recording $dI/dV(x,E)|_y$ as a function of the position $x$ along specific lines in the sample, the feedback loop was closed before moving from one point to the next. $dI/dV(x,y)|_E$ maps at specific energies were measured under closed feedback loop conditions.

As previously reported \cite{Stradi2016,Kamna1998}, TCNQ molecules self-assemble onto the Cu(111) surface with a rhomboidal unit cell of size 4.8 nm $\times$ 1.8 nm (long diagonal $\times$ short diagonal). The islands tend to be elongated along the direction of the short diagonal of the unit cell, and the island edges in this direction follow straight lines with a rather small number of defects (Fig. 1a). The molecular structure at these edges is not simply the crystal termination of the 2D self-assembled structure: comparison between the 2D Fourier Transform (FT) of images of the molecular structure inside the TCNQ island and the 1D FT of a scan line at the edge reveals that the latter presents a new set of peaks halfway between the projection of the reciprocal space of the bulk (1.8 nm$^{-1}$, Fig. 1b). Thus, a 1D reconstruction of the island edge takes place, with a periodicity of 3.5 nm, which is twice the size of the short rhombus diagonal. The 1D unit cell of the reconstructed edges consists of three molecules, one parallel to the edge, and two forming $\pm 36^\circ$ with the edge in the shape of a V (Fig. 1c). STM images recorded at low voltages ($<$100 mV) close to the reconstructed island edge reveal the standing wave pattern caused by interference between incoming and scattered electrons (Fig. 1c). Standing waves surrounding individual molecular adsorbates \cite{Gross2004} or arrays or organic molecules \cite{Pennec2007,Wyrick2011,Lobo-Checa2009} have been previously found, but in previous studies (both for organic adsorbates or other types of 0 and 1D defects \cite{Avouris1994,Crommie1993a,Crommie1993a,Crommie1993b,Hasegawa1993}) the wave fronts follow 1D continuous curves on the surface, instead of a 2D array of maxima and minima like the one in Fig. 1c.

\begin{figure}[t]
\includegraphics[]{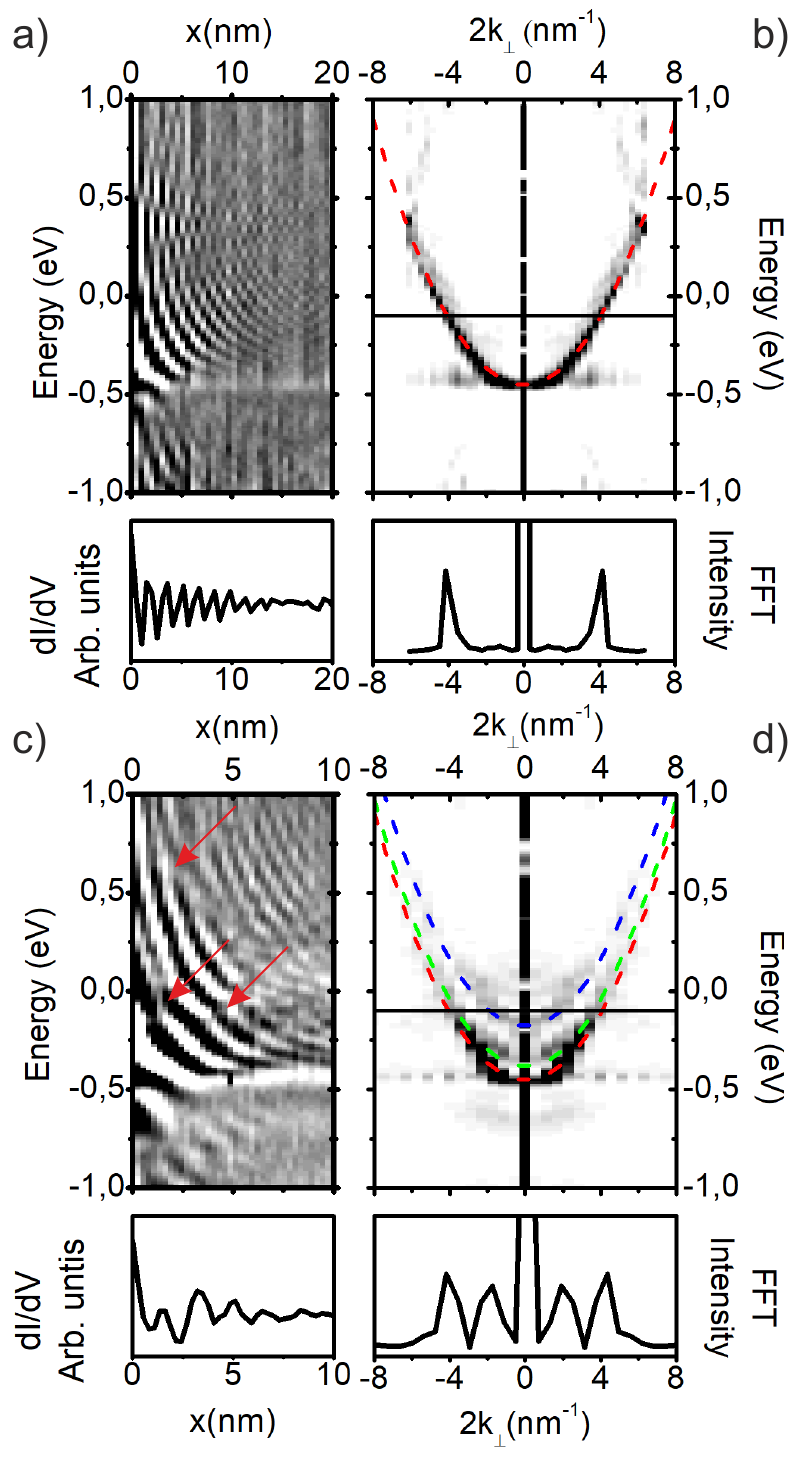}
\caption[]{
\label{figure2}
(a) Upper panel: $dI/dV(x,E)$ measured along the perpendicular direction of a bare step edge. Lower panel: Cut of the upper panel for a bias voltage of -100 mV. (b) Upper panel: FFT of the data of (a), revealing the existence of only one band. Bottom panel: cut of the upper panel for a bias voltage of -100 meV. (c) Upper panel: $dI/dV(x,E)$ measured along the perpendicular to the TCNQ island edge. Lower panel: Cut of the upper panel for a bias voltage of -100 mV. (d) Upper panel: FFT of the data of (b), revealing the existence of three sub-bands. Bottom panel: cut of the upper panel for a bias voltage of -100 meV.}
\end{figure}

In order to investigate the electron scattering processes at the TCNQ island edge leading to this peculiar standing wave pattern, we have recorded $dI/dV(x,E)$ spectra at different distances $x$ from the TCNQ edges (Fig. 2c and d) and from monoatomic step edges on the bare Cu(111) surface (Fig. 2a-b). In both cases, the spectra reveal the onset of conductivity at about 0.45 eV below the Fermi level, characteristic of the Cu(111) surface state. For higher energies, the conductivity oscillates with the distance to the linear scatterer, with a period that decreases with increasing energies (Fig. 2a and c, upper panels). The most striking difference between the scattering at TCNQ edges and at bare Cu(111) steps is the dependence of the amplitude with the distance: while for bare monoatomic steps the intensity decreases monotonically with the distance to the step (Fig. 2a, lower panel), as previously reported \cite{Burgi2000,Hasegawa1993,Crommie1993a}, for TCNQ edges there is an additional amplitude modulation (Fig. 2c, lower panel). Also, while the distance at which a particular wavefront is found for scattering with a copper step edge increases smoothly with decreasing energies, for scattering with TCNQ island edges the evolution present well defined kinks or discontinuities, marked by arrows in Fig. 2c. Both phenomena can be attributed to the superposition of waves with different wavelengths. This is more clearly seen by performing a line-by-line FT of the experimental data at each energy, showing the tunneling conductance as a function of $\Delta k_{\perp}$ and $E$. For scattering with a bare Cu(111) step, only one wave vector is found to contribute at each energy above the surface state onset, with the parabolic dispersion relation of the surface state (solid red line in Fig. 2b). Previous studies demonstrated that the periodicity of the standing wave pattern due to reflection of surface state electrons with bare step edges is determined by scattering processes of electrons in normal incidence, for which $\Delta k_{\perp}=2k_{\perp}$, while the effect of other scattering geometries is a decay in the intensity of the maxima away from the step edge \cite{Burgi2000}. On the other hand, for scattering with the TCNQ edge, alongside the band originating from the normal reflection of surface state electrons, two other parabolas are clearly visible at higher energies (Fig. 2d, upper panel). The bottom of the new bands are shifted with respect to the surface state onset by 70 and 270 meV respectively, but the effective mass does not change significantly (blue and green curves in Fig. 2d). The electronic structure of the 2DEG nearby a single edge of the molecular islands thus consists in a discrete series of electronic sub-bands which disperse parabolically with $k_{\perp}$, even when a second edge is not present to produce confinement.

\begin{figure}[h]
\includegraphics[]{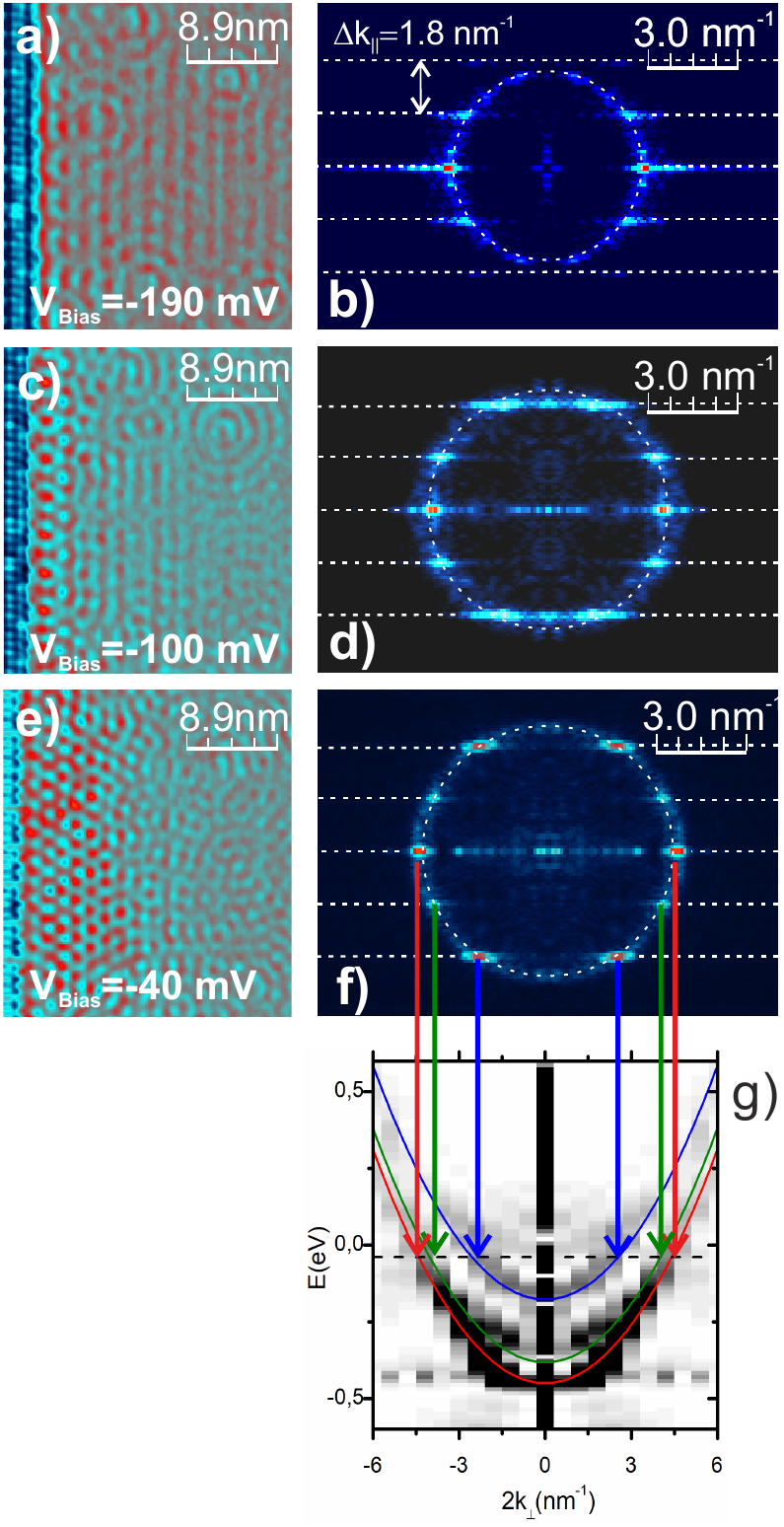}
\caption[]{
\label{figure3}
$dI/dV(x,y)|_E$ maps recorded with the lock-in amplifier at different bias voltages (a, c and e) and their corresponding symmetrized 2D-FFT images (b,d and f). At the constant-energy circles arising from scattering ad different types of surface defects (white dotted circles), some points appear with a rather large intensity, being equally spaced in $\Delta k_{\parallel}$ (white dotted lines). (g) Comparison of the projection onto the $\Delta k_{\perp}$ axis of the high-intensity points in the 2D-FFT with the momenta of the three parabolas found in Fig. 2.
}
\end{figure}

The origin of the discrete sub-band structure has been further examined by recording $dI/dV(x,y)|_E$ maps with the feedback loop closed at specific bias voltages, showing different standing wave patterns (Fig. 3(a), (c) and (e)). The wave-vectors that contribute to the observed patterns have been studied by 2D FT of those images (see Figure 3(b), (d) and (f)): depending on the energy, the FFT of the standing waves shows intense maxima from three or five pairs of points superimposed on a faint circular background attributed to scattering with randomly distributed point-like defects. These maxima disappear gradually with increasing distance to the island edge, so they can be attributed to scattering of surface state electrons with TCNQ molecules. The momentum transfer values parallel to the edge for the observed maxima, $\Delta k_{\parallel}$, are equally spaced by 1.8 nm$^{-1}$ for all the energies investigated. Comparison between the 2D maps and the energy resolved $dI/dV(k_{\perp},E)$ from Figure 2a demonstrates that the new sub-bands observed at the TCNQ edges correspond to the quasi-free dispersion relation of the surface state electrons of Cu(111) with $k_{\perp}$ at the quantized values of $\Delta k_{\parallel,n}$ (Fig. 3g).

\begin{figure*}[t]
\includegraphics[width=0.8\linewidth]{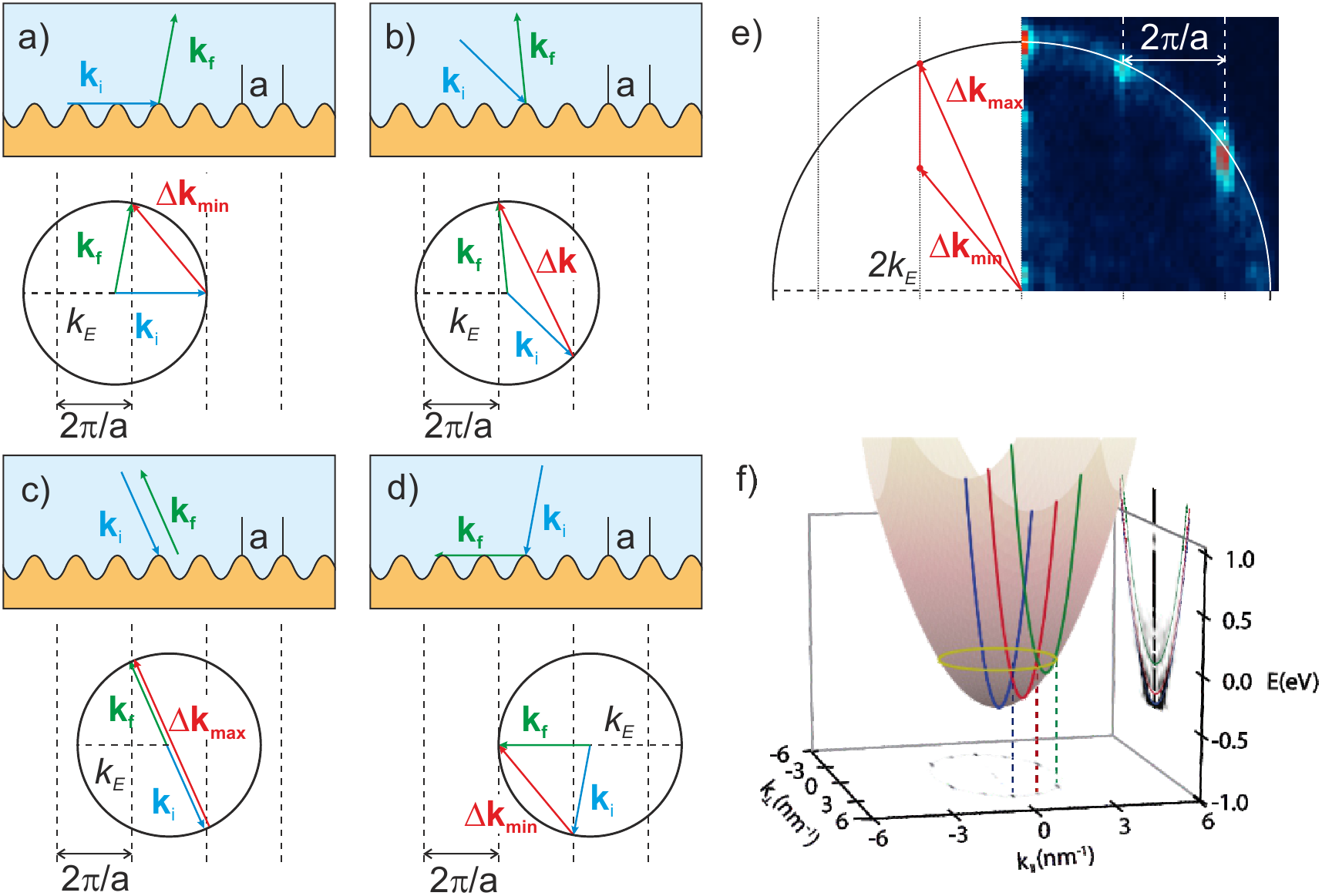}
\caption[]{
\label{figure4}
(a-d) Scattering geometry for four possible incidence angles of the incoming electron, calculated for an energy of -40 meV and $n=-1$ in the Laue condition. Top panel: Real space representation of the scattering process, where the orange area corresponds to the TCNQ edge, and the light cyan area to bare Cu(111) areas. Blue (green) vectors represent the momentum of the incident (scattered) electron. Bottom panels: Reciprocal space representation of the scattering process visualized through an Ewald-sphere construction. Vertical dotted lines correspond to the reciprocal space of the TCNQ edge, while the solid circle represent the constant-energy surface for the surface state electrons. Momentum transfer vectors associated to these scattering processes are also drawn in red. (a) and (d) correspond to the limiting cases in which one of the two momentum vectors is aligned with the island edge. (c) corresponds to the backscattering configuration. (e) Momentum-transfer space derived from the previous analysis compared to the experimental $dI/dV(x,y)|_E$ maps at -40 meV, showing that the vectors contributing to the standing waves are those corresponding to backscattering processes. (f) Origin of the discrete sub-bands as intersections between the 2D surface state paraboloid and the constant $k_{\parallel}$ planes originating from the Laue and backscattering conditions ($k_{\parallel,n}=nG/2$).
}
\end{figure*}

The quantization of the parallel momentum transfer, $\Delta k_{\parallel,n}=nG$, is strongly reminiscent of the Laue condition for diffraction of a 2D wave with a 1D crystal. The reciprocal space of the 1D TCNQ edge consists of a regular array of straight lines perpendicular to the edge with a spacing of $2\pi/a=1.8$ nm$^{-1}$ (see Fig. 1b), in perfect agreement with the value of $G$ within experimental error. From that point of view, our experimental results are comparable to a Low Energy Electron Diffraction experiment in which the probe beam is composed of bound 2D electrons instead of free 3D electrons. One important difference, however, is the direction of the incident electrons: while in standard diffraction techniques the incidence angle is chosen by the experimentalist, in our results surface state electrons will arrive at the TCNQ edge from every possible direction in the half-plane not occupied by the island (Figs. 4a-d, top panel).

The momentum of the outgoing electron after a Bragg scattering event is completely determined by the direction of the incoming electron through the Laue and energy-conservation conditions. This is illustrated in the Ewald sphere construction in the lower panels of Figs. 4a-d for an energy of -40 meV and $n=-1$ in the Laue condition. In this construction, the constant-energy circles are drawn so that the incident momentum vector links the center of the circles with one of the reciprocal space lines of the TCNQ island edge. The outgoing momentum following Bragg diffraction of order $n$ is thus determined from the intersection of the constant-energy circle with another reciprocal space line separated by $nG$ from the one corresponding to the incident momentum. As can be observed in Figure 4, each incidence angle will lead to a different momentum transfer vector ($\Delta \mathbf{k}$). The modulus of $\Delta \mathbf{k}$ is maximum for the incidence angle that makes the initial and final momentum vectors antiparallel (backscattering configuration, $\mathbf{k_i}=-\mathbf{k_f}$, $|k_{\parallel,i}|=|k_{\parallel,f}|=nG/2$ see Fig. 4c) since the momentum transfer vector couples points which are diametrically opposed in the constant energy circle. For each energy the range of possible scattering geometries is limited by the case of grazing incidence of incoming electrons ($k_{\parallel,i}=\pm k_E$, Fig. 4a), and the case of grazing direction for the scattered electron ($k_{\parallel,f}=\mp k_E$, $k_{\parallel,i}=\mp k_E \pm nG$, Fig. 4d). The modulus of $\Delta \mathbf{k}$ has the minimum value for these two limiting scattering configurations. In the momentum transfer space (which is directly comparable to the $dI/dV(x,y)|_E$ maps), n\textsuperscript{th}-order Bragg scattering will be defined by a range of momentum transfer vectors with the same parallel component ($nG$) and a value of the modulus ranging from the maximum of $2k_E$ for backscattering processes to the minimum value for the limiting scattering configurations (Figure 4e). Notice that all these processes are elastic since they couple points in the same constant-energy surface.

Comparing the experimental 2D-FFT with the expected distribution of momentum transfer values (Fig. 4e) we observe that, although some intensity is visible at the reciprocal space lines extending to the interior of the backscattering circle, the intensity is strongly localized at the circle. From our previous discussion, the localization of the intensity at the $2k_E$ circle in the momentum transfer space implies that, out of all the possible scattering geometries, those with the largest contribution to the observed standing wave pattern correspond to the backscattering configuration. This is in agreement with previous studies on standing wave patterns of noble metal surface states observed by STM \cite{Burgi2000,Crommie1993b,Avouris1994,Gross2004,Pennec2007}, and is related to the fact that the momentum transfer vectors associated to these scattering processes are extremal over all the possible vectors spanning the constant energy lines. Notice that taking into consideration both the Laue condition for Bragg diffraction ($\Delta k_{\parallel,n}=k_{\parallel,f}-k_{\parallel,i}=nG$) and the backscattering condition ($k_{\parallel,i}=-k_{\parallel,f}$) we must conclude that the DOS of the sample is dominated by electrons for which the parallel component of both the incoming and outgoing momenta is quantized in units of $G/2$. The electronic structure of the Cu(111) surface state nearby the edges of TCNQ islands is thus discretized by this process, and the original surface state becomes a discrete collection of sub-bands obtained from the intersection of the 2D paraboloid with a set of equally spaced planes of constant $k_{\parallel}=nG/2$ (see Fig. 4f). The bottoms of these sub-bands are shifted upwards in energy by $\hbar^2/2m^*(nG/2)^2=\hbar^2/2m^*(n\pi/a)^2$. These shifts become very large for a short periodicity of the scatterers. For example, for the bottom of the 1\textsuperscript{st} sub-band to be 1 eV above the Fermi level, the periodicity must be of at least 0.8 nm. Thus, our observation is only possible because of the large periodicity of the TCNQ edge.

To summarize, the diffraction of 2D electrons with a 1D periodic array of molecular scatterers leads to the quantization of the momentum transfer parallel to the edge in integer units of the reciprocal space vector of the edge $G$ and, thus leads to the existence of discrete sub-bands without full quantum confinement at the corrugated edges of TCNQ islands. In general, such scenario should discretize the electronic structure of solids close to periodically corrugated interfaces, creating a layer of thickness of the order of the electronic coherence length with a modified DOS. Engineering the lateral periodicity of the surface corrugation becomes thus a new avenue to modify the physical properties of materials, from the superconducting transition temperature to their thermal stability.

\begin{acknowledgments}
The authors acknowledge financial support from the Spanish Ministry for Economy and Competitiveness (grants FIS2015-67367-C2-1-P, FIS2015-72482-EXP), the regional government of Comunidad de Madrid (grants S2013/MIT-3007 and S2013/MIT-2850), Universidad Aut\'onoma de Madrid (UAM/48) and IMDEA Nanoscience.
\end{acknowledgments}

\end{document}